\documentclass[preprint,notitlepage,showpacs,showkeys,preprintnumbers,amsmath,amssymb,superscriptaddress,nofootinbib]{revtex4-1}

\usepackage{graphicx}
\usepackage{dcolumn}
\usepackage{bm}
\usepackage{natbib}
\usepackage[usenames,dvipsnames]{xcolor}
\usepackage{pifont}
\usepackage{relsize}
\definecolor{lightgreen}{rgb}{0,1,0}
\definecolor{darkgray}{gray}{0.20}
\usepackage{subfigure}
\usepackage{hyperref}
\usepackage{float} 
\usepackage{multirow} 
\usepackage{afterpage} 

\usepackage{epstopdf}

\renewcommand\thetable{\arabic{table}} 

\begin{document}

\title{Global Value Trees}

\author{Zhen Zhu} \altaffiliation{Corresponding author. Email: \href{mailto:zhen.zhu@imtlucca.it}{zhen.zhu@imtlucca.it}.} \affiliation{IMT Institute for Advanced Studies Lucca, Lucca, Italy} 
\author{Michelangelo Puliga} \affiliation{IMT Institute for Advanced Studies Lucca, Lucca, Italy}\affiliation{Linkalab, Complex Systems Computational Laboratory, Cagliari, Italy} 
\author{Federica Cerina} \affiliation{Linkalab, Complex Systems Computational Laboratory, Cagliari, Italy} \affiliation{Department of Physics, Universit\`{a} degli Studi di Cagliari, Cagliari, Italy}
\author{Alessandro Chessa}  \affiliation{IMT Institute for Advanced Studies Lucca, Lucca, Italy} \affiliation{Linkalab, Complex Systems Computational Laboratory, Cagliari, Italy}     
\author{Massimo Riccaboni} \affiliation{IMT Institute for Advanced Studies Lucca, Lucca, Italy} \affiliation{Department of Managerial Economics, Strategy and Innovation, Katholieke Universiteit Leuven, Leuven, Belgium}

\begin{abstract}
\noindent The fragmentation of production across countries has become an important feature of the globalization in recent decades and is often conceptualized by the term, global value chains (GVCs). When empirically investigating the GVCs, previous studies are mainly interested in knowing how global the GVCs are rather than how the GVCs look like. From a complex networks perspective, we use the World Input-Output Database (WIOD) to study the global production system. We find that the industry-level GVCs are indeed not chain-like but are better characterized by the tree topology. Hence, we compute the global value trees (GVTs) for all the industries available in the WIOD. Moreover, we compute an industry importance measure based on the GVTs and compare it with other network centrality measures. Finally, we discuss some future applications of the GVTs. 
\end{abstract}




\maketitle

\section{Introduction} \label{sec:intro}

The history of globalization has been marked by two great unbundlings, the first being the spatial separation of production and consumption (i.e., international trade in final products), and more recently, the second being the spatial fragmentation within production (i.e., international trade in tasks and supply chains) \cite{baldwin2006globalisation,baldwin2014supply}. The second great unbundling is often conceptualized by the term, global value chains, or GVCs (Other similar concepts used in the literature include global supply chains \cite{costinot2013elementary}, supply-chain trade \cite{baldwin2014supply}, international fragmentation \cite{los2014global}, outsourcing \cite{feenstra1999impact}, offshoring \cite{grossman2008trading}, and vertical specialization \cite{hummels2001nature}.), since it captures the fact that the value-added of a final product can be distributed globally. In other words, a product (and its components) may have crossed multiple country borders before it arrives in a final consumer's hands. For instance, before it hits the US market, an Apple's iPod needs to be assembled in China, which in turn sources microchips and software from Japan, South Korea, and the US itself \cite{Dedrick2010who}.   

Quite a few theoretical models have been developed to understand the GVCs' structure, mechanism, welfare impacts, and policy implications \cite{grossman2008trading,baldwin2013spiders,costinot2013elementary}. Thanks to the recently constructed global multi-regional input-output (MRIO) tables, empirical studies can be conducted at the industry level and hence identify a more general pattern of the GVCs than do the case studies on the specific products such as iPod. In particular, the global value-added content of exports for a given industry or country can be measured \cite{feenstra1999impact,hummels2001nature,johnson2012accounting,baldwin2014supply,johnson2014five,los2014global,timmer2014slicing}.

Although previous studies can tell us how global the GVCs are, very little is known about how the GVCs look like (A fairly comprehensive survey of the GVCs literature is conducted by Amador and Cabral \cite{amador2014global}. There are a number of studies exploring some structural properties of the GVCs such as the length of a GVC and the industry upstreamness with respect to final consumption \cite{antras2012measuring,backer2013mapping}. However, they only provide some rough estimates of the structural properties rather than any topological details of the GVCs.). To fill the gap in the literature, our paper is the first attempt to investigate the topological properties of the industry-level GVCs. From a complex networks perspective, we map the World Input-Output Database (WIOD) into the global value networks (GVNs), where the nodes are the individual industries in different countries and the edges are the value-added contribution relationships. 

Based on the GVNs, this paper makes some significant contributions to the literature of the GVCs. First, unlike the previous literature which provides only some rough estimates of the structural properties of the GVCs, we are able to produce a detailed topological view of the industry-level GVCs. We compute the global value trees (GVTs) for all the industries available in the WIOD by a breadth-first search algorithm with a threshold of edge weight and a limit of the number of rounds. We explore some basic properties of the GVTs. In particular, we estimate the allometric scaling exponents and find that the GVTs are topologically more similar to a star than to a chain. However, the GVTs have become more and more hierarchical over time. Second, we develop an industry importance measure based on the GVTs and compare it with other network centrality measures of the industries. We find that the tree-based measure performs the best in terms of the correlation with the industry total value-added. Therefore, the GVTs still retain the essential information of the GVNs and can be viewed as a reasonable simplification of the latter. Third, with the rich topological information, the GVTs enable a broad range of empirical studies of the global fragmentation of production such as to examine the evolution of the GVTs for a certain industry and to compare the GVTs of the same industry in different countries.  

The rest of the paper is structured as follows. Section \ref{sec:method} maps the WIOD database into the GVNs and develops an algorithm to compute the GVTs. Section \ref{sec:results} explores some basic properties of the GVTs. In particular, we quantify the allometric scaling pattern of the GVTs and propose an industry importance measure based on the GVTs and compare it with other network centrality measures. Finally, Section \ref{sec:discussion} discusses some future applications of the GVTs and concludes the paper.

\section{Methods} \label{sec:method}

The complex networks approach has been widely used in economics and finance in recent years \cite{kitsak2010,pammolli2002,riccaboni2010,riccaboni2013,chessa2013,nature,zhu2014rise,cerina2014world}. Designed to keep track of the inter-industrial relationships, the input-output system is an ideal test bed for network science. In particular, the global MRIO system can be viewed as an interdependent complex network, where the nodes are the individual industries in different countries and the edges are the input-output relationships between industries \cite{cerina2014world}.

This paper takes one step further and uses the WIOD database to construct the global value networks (GVNs), where the nodes are the individual industries in different countries and the edges are the value-added contribution relationships (The call for a network analysis of the GVCs has existed for years \cite{sturgeon2001we,coe2004globalizing,coe2008global,sturgeon2008value}.). Moreover, based on the GVNs, the global value trees (GVTs) can be computed in a straightforward manner.

\subsection{Data Description} \label{subsec:data}

We use the World Input-Output Database (WIOD) \cite{timmer2012world} to compute the GVNs and the GVTs. At the time of writing, the WIOD input-output tables cover 35 industries for each of the 40 economies (27 EU countries and 13 major economies in other regions) plus the rest of the world (RoW) and the years from 1995 to 2011 (Tables S1 and S2 in Supplementary Information have the lists of countries and industries covered in the WIOD.). For each year, there is a harmonized global level input-output table recording the input-output relationships between any pair of industries in any pair of economies. The numbers in the WIOD are in current basic (producers') prices and are expressed in millions of US dollars (The basic prices are also called the producers' prices, which represent the amount receivable by the producers. An alternative is the purchases' prices, which represent the amount paid by the purchases and often include trade and transport margins. The former is preferred by the WIOD because it better reflects the cost structures underlying the industries \cite{timmer2012world}.). Table \ref{table_1} shows an example of a global MRIO table with two economies and two industries. The $4\times4$ inter-industry table is called the transactions matrix and is often denoted by $\boldsymbol{\mathrm{Z}}$. The rows of $\boldsymbol{\mathrm{Z}}$ record the distributions of the industry outputs throughout the two economies while the columns of $\boldsymbol{\mathrm{Z}}$ record the composition of inputs required by each industry. Notice that in this example all the industries buy inputs from themselves, which is often observed in real data. Besides intermediate industry use, the remaining outputs are absorbed by the additional columns of final demand, which includes household consumption, government expenditure, and so forth (In Table \ref{table_1} we only show the aggregated final demand for the two economies.). Similarly, production necessitates not only inter-industry transactions but also labor, management, depreciation of capital, and taxes, which are summarized as the additional row of value-added. The final demand matrix is often denoted by $\boldsymbol{\mathrm{F}}$ and the value-added vector is often denoted by $\boldsymbol{\mathrm{v}}$. Finally, the last row and the last column record the total industry outputs and its vector is denoted by $\boldsymbol{\mathrm{x}}$. 
\bigskip
\\
{\bf Insert Table 1 here.}   
\bigskip 

\subsection{Construct the Global Value Networks} \label{subsec:GVNs}

If we use $\boldsymbol{\mathrm{i}}$ to denote a summation vector of conformable size, i.e., a vector of all 1's with the length conformable to the multiplying matrix, and let $\boldsymbol{\mathrm{F}}\boldsymbol{\mathrm{i}}=\boldsymbol{\mathrm{f}}$, we then have $\boldsymbol{\mathrm{Z}}\boldsymbol{\mathrm{i}}+\boldsymbol{\mathrm{f}}=\boldsymbol{\mathrm{x}}$. Furthermore, if dividing each column of $\boldsymbol{\mathrm{Z}}$ by its corresponding total output in $\boldsymbol{\mathrm{x}}$, we get the so-called technical coefficients matrix $\boldsymbol{\mathrm{A}}$ (The ratios are called technical coefficients because they represent the technologies employed by the industries to transform inputs into outputs.). Replacing $\boldsymbol{\mathrm{Z}}\boldsymbol{\mathrm{i}}$ with $\boldsymbol{\mathrm{A}}\boldsymbol{\mathrm{x}}$, we rewrite the above equation as $\boldsymbol{\mathrm{A}}\boldsymbol{\mathrm{x}}+\boldsymbol{\mathrm{f}}=\boldsymbol{\mathrm{x}}$. It can be rearranged as $(\boldsymbol{\mathrm{I}}-\boldsymbol{\mathrm{A}})\boldsymbol{\mathrm{x}}=\boldsymbol{\mathrm{f}}$. Then we can solve $\boldsymbol{\mathrm{x}}$ as follows:
\begin{equation}
\boldsymbol{\mathrm{x}}=(\boldsymbol{\mathrm{I}}-\boldsymbol{\mathrm{A}})^{-1}\boldsymbol{\mathrm{f}}
\end{equation}
where matrix $(\boldsymbol{\mathrm{I}}-\boldsymbol{\mathrm{A}})^{-1}$ is often denoted by $\boldsymbol{\mathrm{L}}$ and is called the Leontief inverse \cite{leontief1936quantitative,miller2009input}.

If dividing each element of $\boldsymbol{\mathrm{v}}$ by its corresponding total output in $\boldsymbol{\mathrm{x}}$, we get the value-added share vector and denote it by $\boldsymbol{\mathrm{u}}$. Moreover, if we use $\boldsymbol{\mathrm{\hat u}}$ to denote a diagonal matrix with $\boldsymbol{\mathrm{u}}$ on its diagonal, then the value-added contribution matrix can be computed as follows:
\begin{equation}
\boldsymbol{\mathrm{G}}=\boldsymbol{\mathrm{\hat u}}\boldsymbol{\mathrm{L}}
\end{equation}
where $\boldsymbol{\mathrm{G}}$ is the value-added contribution matrix and its element $0\leq G_{ij}\leq 1$ is industry $i$'s share of the value-added contribution in industry $j$'s final demand, $f_j$. 

Finally, the GVNs can be constructed by using $\boldsymbol{\mathrm{G}}$ as the adjacency matrix. Notice that the GVNs are both directed and weighted (We don't consider the self-loops so that we replace the diagonal of $\boldsymbol{\mathrm{G}}$ with zeros. Meanwhile, we don't consider the rest of the world (RoW) and focus our attention on the 40 countries available in the WIOD.).

\subsection{Compute the Global Value Trees} \label{subsec:GVTs}

Based on the GVNs, the GVTs can be obtained by a modified breadth-first search algorithm. First, we choose an industry as the root of the GVT and the tree grows as we add the most relevant industries to the root industry in terms of the value-added contribution. Second, since the GVNs are almost completely connected (This is a general feature of the input-output networks due to the aggregated industry classification \cite{cerina2014world}.), we search the GVTs based on a threshold of the edge weight, which we denote by $\alpha$, in order to separate the most relevant industries from the less relevant ones. Third, we limit the breadth-first search to a fixed number of rounds, which we denote by $\gamma$. Again, this is to ensure that only the most relevant industries with respect to the root industry are included in the GVTs.

Our benchmark GVTs are based on $\alpha=0.01$ and $\gamma=3$. The tree topology requires that $\gamma\geq 2$ because it would rather become a star topology if $\gamma=1$. We choose $\gamma=3$ to ensure that the nodes included in the GVTs are economically relevant to the root industries. To choose a proper value of $\alpha$, we gather some statistics of the number of nodes across the GVTs by holding constant $\gamma=3$ and by only varying the value of $\alpha$. Table \ref{table_2} has the summary statistics of the size of the GVTs based on $\alpha=0.1$, $\alpha=0.01$, and $\alpha=0.001$ and for the selected years 1995, 2003, and 2011, respectively. 

We choose $\alpha=0.01$ for a number of reasons. First, most of the industries in the WIOD have the corresponding GVTs if $\alpha=0.01$. The number of observations, i.e., nonempty GVTs, under $\alpha=0.01$ is just slightly smaller than under $\alpha=0.001$. In contrast, out of the 1400 industries, fewer than 200 GVTs can be computed if the threshold is set as high as 0.1. Second, $\alpha=0.01$ provides us with much more manageable size of GVTs (around 40 nodes on average) than $\alpha=0.001$ (around 800 nodes on average). In other words, only the most relevant nodes to the root industry will be present in the GVTs if $\alpha=0.01$. Last but not least, the coefficient of variation is the highest when $\alpha=0.01$, which means that $\alpha=0.01$ provides us with a more diverse set of GVTs than the other two parameter choices.  This is very helpful if we want to examine the different topological properties of the different GVTs.  
\bigskip
\\
{\bf Insert Table 2 here.}   
\bigskip

\begin{figure}[H]
\centering
{\includegraphics[width=16cm]{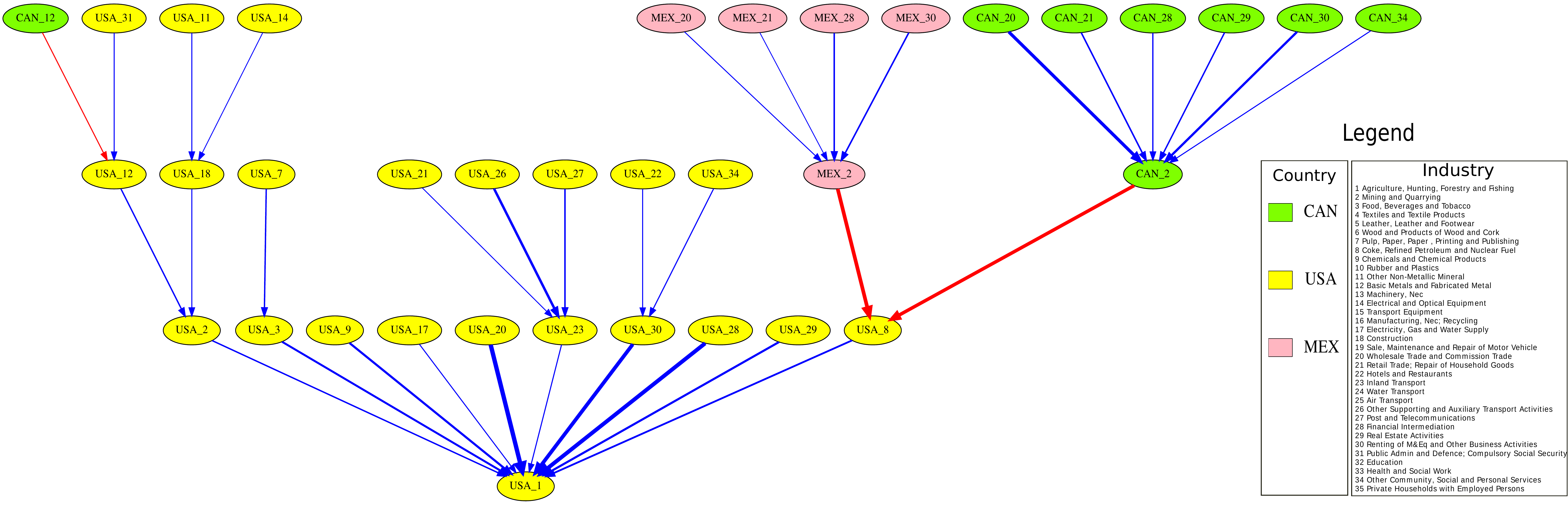}}
\caption{{\bf The GVT of USA's agriculture industry in 2011.} The edge weight threshold is set to 0.01 and the number of rounds is limited to 3. Different colors of the nodes indicate different countries. The red edges indicate cross-country relationships while the blue edges indicate domestic relationships. The edge width is proportional to the edge weight, i.e., the share of the value-added contribution.} \label{fig_1}
\end{figure}

Figure \ref{fig_1} shows the GVT of USA's agriculture industry in 2011, with $\alpha=0.01$ and $\gamma=3$. The economic interpretation is that, if the final demand in the root is one million US dollars, then all the industries contributing more than 1\% locally (to the direct neighbors) and directly or indirectly contributing more than 1 dollar (one million multiplied by ${0.01}^3$) to the root are included in the tree.  
\bigskip
\\
{\bf Insert Figure 1 here.}   
\bigskip

\section{Results} \label{sec:results}

Once we have computed the GVTs, some basic properties of the tree topology can be explored. Subsection \ref{subsec:allometry} quantifies the allometric scaling pattern of the GVTs. We estimate the allometric scaling exponents and find that the GVTs are topologically more similar to a star than to a chain. However, the GVTs have become more and more hierarchical over time. Subsection \ref{subsec:measure} proposes a tree-based industry importance measure and compares it with other network centrality measures. We find that the tree-based measure performs the best in terms of the correlation with the industry total value-added. Therefore, the GVTs still retain the essential information of the GVNs and can be viewed as a reasonable simplification of the latter.

\subsection{Allometric Scaling Pattern} \label{subsec:allometry}

The allometric scaling pattern refers to the power law relationship between size and other physical or behavioral variables. Previous studies have documented the ubiquitous existence of the allometric scaling pattern in systems as diverse as river networks, cellular metabolism, population dynamics, and food web \cite{banavar1999size,garlaschelli2003universal}. 

For a directed tree topology, if we denote the total number of nodes in the sub-tree rooted at node $i$ by $X_i$ and the sum of all $X_i$'s in the sub-tree rooted at node $i$ by $Y_i$, then an allometric scaling relationship is observed between $Y_i$ and $X_i$ and can be described by a power law, i.e., $Y_i\sim X_i^\eta$, where $\eta$ is called the allometric scaling exponent.

\begin{figure}[H]
\centering
{\includegraphics[width=12cm]{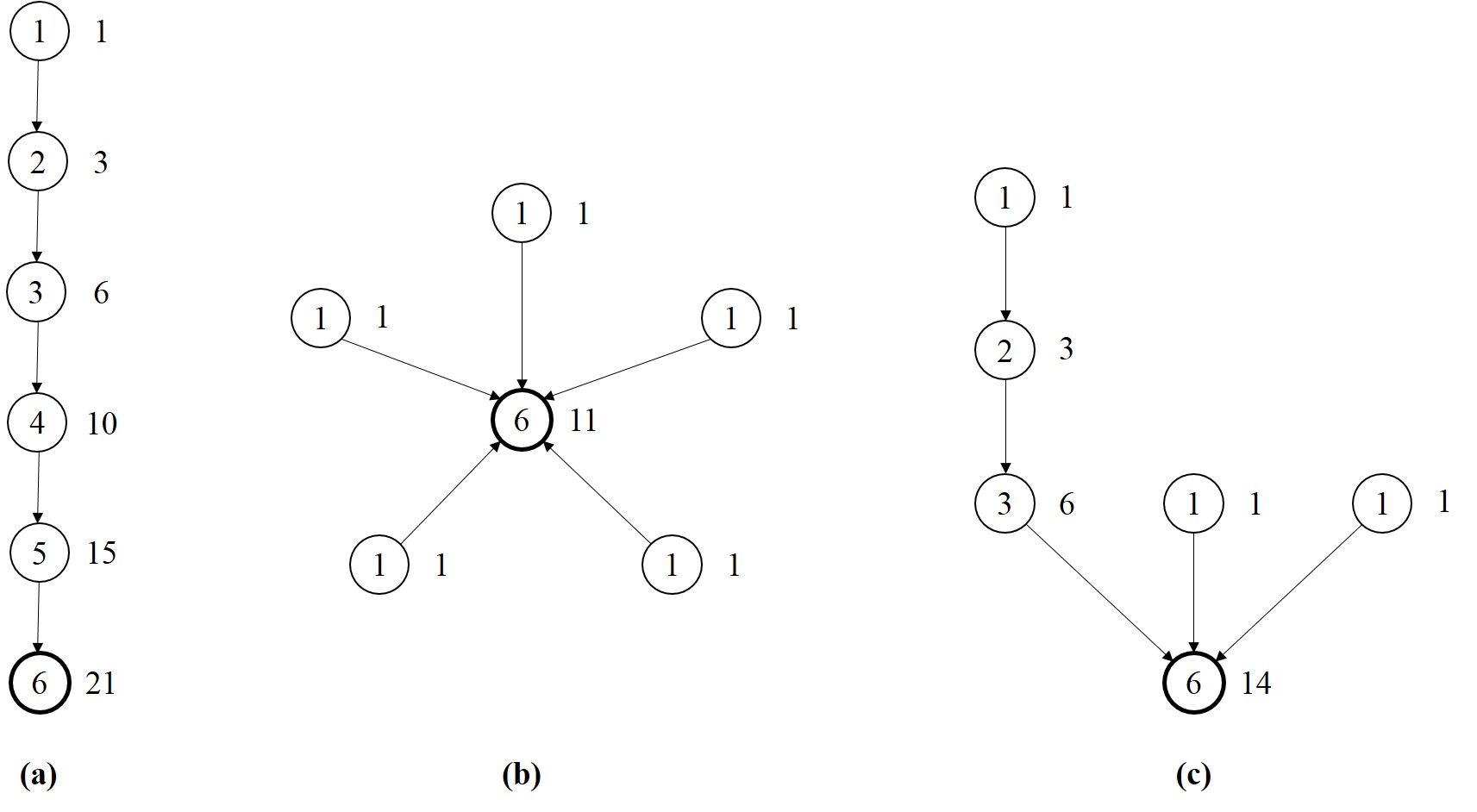}}
\caption{{\bf Examples of the allometric scaling relationship.} The numbers inside the node circles are $X_i$'s whereas those next to the circles are $Y_i$'s. The node with the thick circle is the root. From left to right, they are a chain (a), a star (b), and a tree (c), respectively.} \label{AlloExample}
\end{figure}

Figure \ref{AlloExample} shows the examples of a chain, a star, and a tree, respectively. The numbers inside the node circles are $X_i$'s whereas those next to the circles are $Y_i$'s. The allometric scaling exponent $\eta$ of a tree is lower-bounded by that of a star ($\eta=1$) and upper-bounded by that of a chain ($\eta=2$). As a result, $\eta$ can be interpreted as a measure of hierarchicality, as star is the ``flattest'' topology and chain is the most hierarchical topology given the same number of nodes.  
\bigskip
\\
{\bf Insert Figure 2 here.}   
\bigskip

\begin{figure}[H]
  \centering
{\includegraphics[width=16cm]{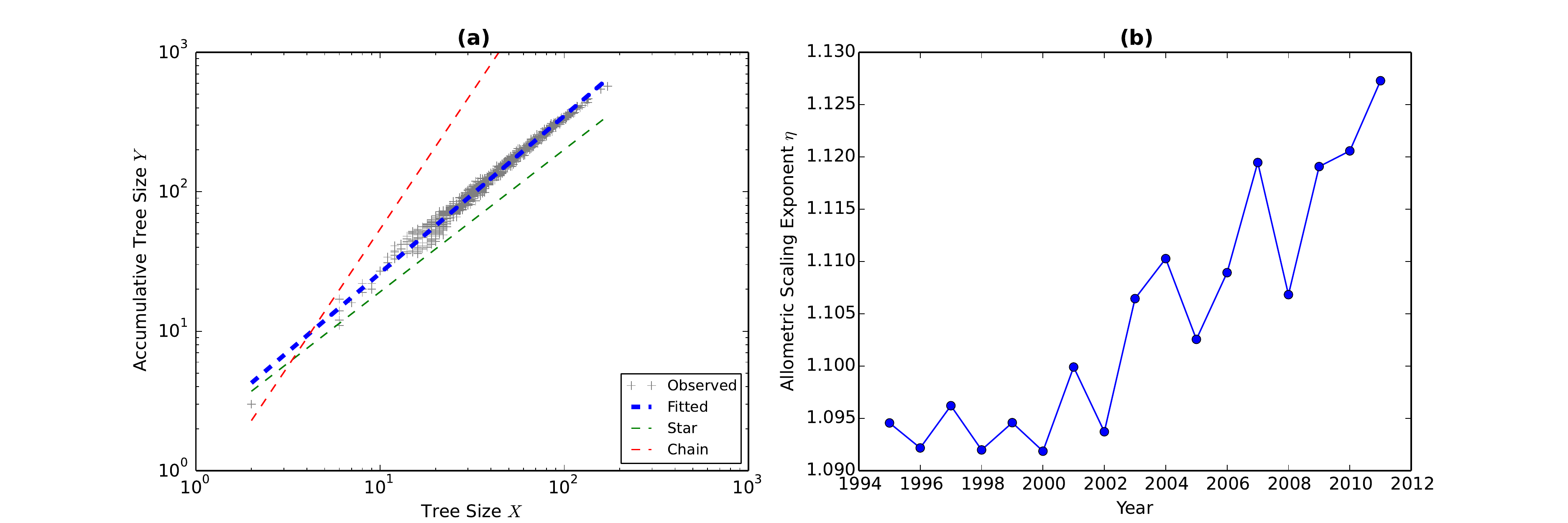}}
  \caption{{\bf Estimation of the allometric scaling exponent $\eta$.} Panel (a) shows the log-log plot of the root-node $Y_i$-$X_i$ pairs in 2011, where the horizontal axis is the $X_i$ of the root node, i.e., the total number of nodes in a given GVT (the tree size), and the vertical axis is the $Y_i$ of the root node, which we call the accumulative tree size. The gray crosses are the observed data points. The thick blue dashed line is fitted with the observed data and with the slope of $\eta$. The fitting lines for star and chain based on the same set of $X_i$'s are the green dashed line and the red dashed line respectively. Panel (b) plots the estimated $\eta$'s over time.}
  \label{FitLine}
\end{figure}

To examine the hierarchicality of the GVTs, we estimate $\eta$'s based on the root-node $Y_i$-$X_i$ pairs across all the GVTs for each year. Figure \ref{FitLine} has the estimation result of $\eta$. Panel (a) shows the log-log plot of the root-node $Y_i$-$X_i$ pairs in 2011, where the horizontal axis is the $X_i$ of the root node, i.e., the total number of nodes in a given GVT (the tree size), and the vertical axis is the $Y_i$ of the root node, which we call the accumulative tree size. The gray crosses are the observed data points. The thick blue dashed line is fitted with the observed data and with the slope of $\eta$. The fitting lines for star and chain based on the same set of $X_i$'s are the green dashed line and the red dashed line respectively. It is straightforward to see that in 2011 the GVTs are more similar to a star than to a chain. Panel (b) plots the estimated $\eta$'s over time. Again, the values of $\eta$ are all closer to 1 than to 2. However, there is a clear upward trend, which means that the GVTs have become more and more hierarchical over time (Shi et al. \cite{shi2014hierarchicality} also estimate the allometric scaling exponent to understand the hierarchicality of the global production system. However, they consider the directed tree as a flow network. Furthermore, their paper differs from ours in both data source and research strategy. They use the United Nations COMTRADE database to construct the product-specific trade networks while we use the WIOD database to construct the GVNs with both country and industry dimensions.). 
\bigskip
\\
{\bf Insert Figure 3 here.}   
\bigskip

\subsection{A Tree-Based Importance Measure} \label{subsec:measure}

The GVTs are the subgraphs of the GVNs. Unlike the GVNs, the GVTs reveal the local importance of the industries. Previous studies have shown that the subgraph centrality measure can be used to complement the global centrality measures \cite{estrada2005subgraph}. Hence, we compute a simple industry importance measure based on the GVTs and compare it with other network centrality measures. 

First, we denote a tree with the root $r$ by $T(r)$. Furthermore, we denote the total number of nodes in the sub-tree rooted at industry $i$ by $X_i(r)$ and the total number of nodes in the tree $T(r)$ by $N(r)$. If industry $i$ is present in $k$ trees all over the world and we denote the set of roots of the $k$ trees by $S_i$, then the importance of industry $i$ is defined as follows: 
\begin{equation}
TI_i=\sum_{r\in S_i, r\neq i}\frac{X_i(r)}{N(r)}\frac{FD(r)}{WGDP}
\end{equation}
where $TI_i$ is the tree-based importance measure of industry $i$, $FD(r)$ is the final demand in the root industry $r$ and $WGDP$ is the world GDP. Notice that when calculating $TI_i$, we don't consider the role played by industry $i$ in its own GVT (i.e., $r\neq i$), although the input-output network has strong self-loops \cite{cerina2014world}. 

The economic interpretation of the importance measure is that, more important industries are more closely attached to the root and are able to ``pull'' a larger portion of the GVTs (measured by $\frac{X_i(r)}{N(r)}$) and are associated with more important roots (measured by $\frac{FD(r)}{WGDP}$). 

\begin{figure}[H]
\centering
{\includegraphics[width=16cm]{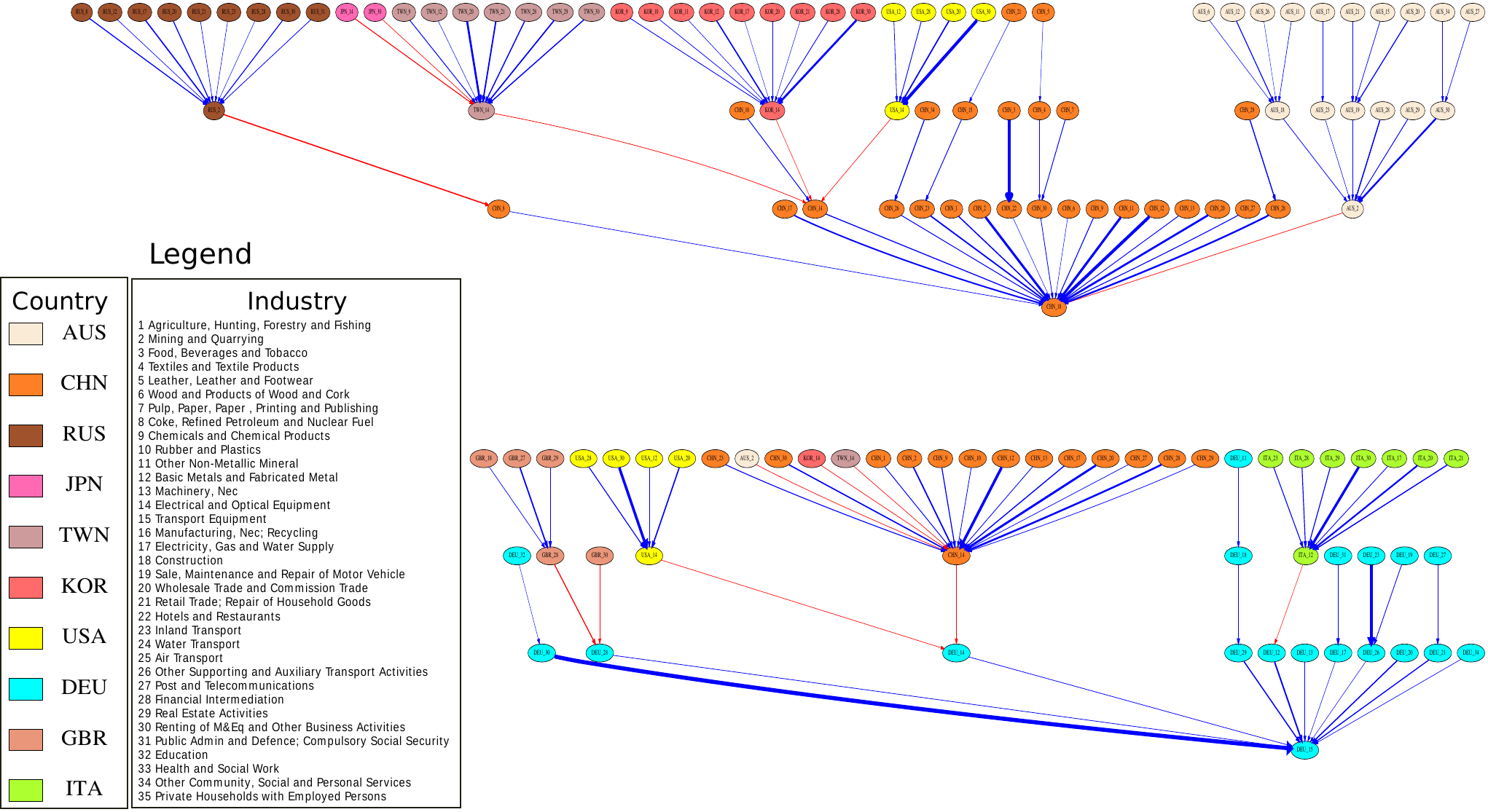}}
\caption{{\bf The domestic and foreign GVTs where China's electrical equipment industry has the highest importance score in 2011.} The upper and lower trees are the domestic and foreign trees respectively where China's electrical equipment industry has the highest importance score. The edge weight threshold is set to 0.01 and the number of rounds is limited to 3. Different colors of the nodes indicate different countries. The red edges indicate cross-country relationships while the blue edges indicate domestic relationships. The edge width is proportional to the edge weight, i.e., the share of the value-added contribution.} \label{fig_2}
\end{figure}

Moreover, since each $T(r)$ where industry $i$ is present has a score of importance, i.e.,$\frac{X_i(r)}{N(r)}\frac{FD(r)}{WGDP}$, we can identify the GVTs where industry $i$ has the highest importance score. For instance, Figure \ref{fig_2} shows the GVTs where China's electrical equipment industry has the highest importance score for domestic and foreign roots respectively in 2011.
\bigskip
\\
{\bf Insert Figure 4 here.}   
\bigskip

To examine the tree-based importance measure in a more systematic way, we compare it with other network centrality measures. Table \ref{table_3} has the top-20 industries identified by different measures for the selected years. Again, $TI$ is the tree-based importance measure. We also provide the results based on some network centrality measures. In particular, $CC$ is the closeness centrality, $BC$ is the betweenness centrality, $PR$ is the PageRank centrality. Finally, we include the measure of economic size of the industries, the industry total value-added, which is denoted by $VT$. Some interesting patterns can be seen from this table. First, all the measures have captured the rise of China over time. Back in 1995, almost no industries from China are identified as the top-20 (The only exception is that China's construction industry takes the 19th place according to the PageRank centrality.). In 2003, China's industries start to be picked up by $TI$, $BC$, and $PR$, which are more topologically sensitive than $CC$ and $VT$. In 2011, China's industries show up in all the measures. Second, each measure captures different information, at least at the top-20 level. In 1995 and 2003, $VT$ is dominated by the service industries in big and advanced economies. But in 2011, the rise of China also brings the industries of agriculture and basic metals to the top list. $CC$ captures big industries in big economies (As shown later in Table \ref{table_4}, $log(CC)$ and $log(VT)$ are strongly correlated.). In 1995 $CC$ is mixed by USA and Germany while it is dominated by USA and China in 2003 and 2011 respectively. $BC$ has a more diverse list of industries in terms of economic size. Besides the big ones, it also includes smaller industries such as Turkey's textiles industry and Indonesia's mining industry. $PR$ is also diverse in terms of economic size. But it is quite stable over time, i.e., except for the rise of China, more or less the same industries are identified as the top-20. Last but not least, $TI$ generally identifies the big industries. But it also gives credits to industries such as Russia's mining industry and Korea's electrical equipment industry, which have strong presence in the GVTs (In Table S3 in Supplementary Information, we also report the country rankings by summing up the measures of the industries in the same country.).    
\bigskip
\\
{\bf Insert Table 3 here.}   
\bigskip

Moreover, Table \ref{table_4} reports the Pearson correlation coefficients among them (in logarithm) for the selected years. For a given year, all the coefficients are based on a common sample among the different measures. It turns out that all the coefficients are positive and almost all of them are significant at 1\% level (The only exceptions are between $log(BC)$ and $log(CC)$ in 2003 and 2011.). 

We find that $TI$ performs the best in terms of the correlation with $VT$. Nevertheless, this is not to say that we should abandon other measures and solely use $TI$ to understand the importance of a given industry. After all, we only consider the intermediate value-added flows when calculating $CC$, $BC$, and $PR$, whereas we also take into account the final demand in the root industry, i.e., $FD(r)$, when calculating $TI$, which gives more power to $TI$ in explaining $VT$. However, the strong correlation between $log(TI)$ and $log(VT)$ at least shows that the GVTs retain the essential information of the GVNs and can be viewed as a reasonable simplification of the latter. That is, $TI$ can be considered as a measure of industry's position advantage. An industry holds an advantageous position by either attaching to big industries (i.e., big $\frac{FD(r)}{WGDP}$) or by affecting big portion of the GVTs (i.e., big $\frac{X_i(r)}{N(r)}$). As a result, the better-positioned industries are more competitive in the world production system and hence are able to extract more value-added across the GVTs. Moreover, since the component $\frac{X_i(r)}{N(r)}$ of $TI$ measures how closely the given industry is attached to the roots (i.e., bigger $\frac{X_i(r)}{N(r)}$ implies smaller distance to the roots), it can be considered as a measure of downstreamness. That is, the higher $TI$ is the more downstream the industry is in the GVTs. Therefore, the strong correlation between $log(TI)$ and $log(VT)$ supports Stan Shih's theory of ``smiling curve'', which states that most value-added potentials are concentrated at the beginning (upstream) and the ending (downstream) parts of the supply chains.  
\bigskip
\\
{\bf Insert Table 4 here.}   
\bigskip

\section{Discussion} \label{sec:discussion}

\begin{figure}[H]
\centering
{\includegraphics[width=16cm]{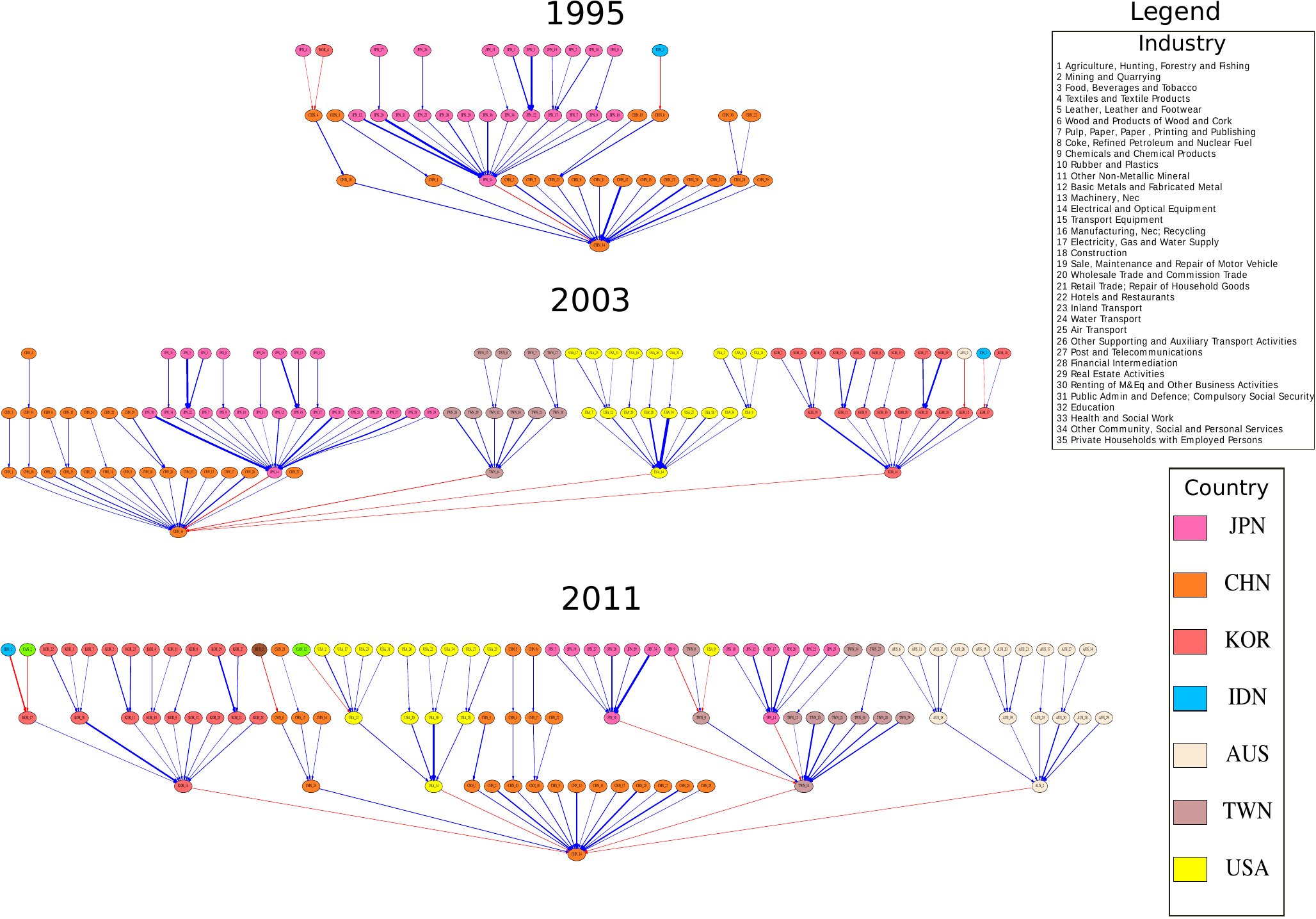}}
\caption{{\bf The evolution of the GVTs rooted at China's electrical equipment industry.} From top down, the GVTs are for 1995, 2003, and 2011 respectively. The edge weight threshold is set to 0.01 and the number of rounds is limited to 3. Different colors of the nodes indicate different countries. The red edges indicate cross-country relationships while the blue edges indicate domestic relationships. The edge width is proportional to the edge weight, i.e., the share of the value-added contribution.} \label{fig_3}
\end{figure}

Once we have the GVTs computed for all the industries available in the WIOD, many interesting questions can be proposed and answered. For instance, does a tree with a fixed root grow over time? This question can be answered by fixing the root industry and examining the GVTs over time. As an example, Figure \ref{fig_3} shows the evolution of the GVTs rooted at China's electrical equipment industry over time. A simple way of measuring the growth of the trees is to count the number of nodes over time. In Figure \ref{fig_3}, the GVT of China's electrical equipment industry evolves from 47 nodes in 1995 to 100 nodes in 2003 and to 106 in 2011. There are also some interesting structural changes in this example. First, Australia's mining industry (AUS\_2) becomes directly attached to the root in 2011. Second, Japan's electrical equipment industry (JPN\_14) is a direct neighbor of the root in 2003 but has to go through Taiwan's electrical equipment industry (TWN\_14) in 2011. The structural changes may have important implications for the industries' competitiveness. In particular, TWN\_14 has become more competitive than JPN\_14 with respect to China's electrical equipment industry based on the $TI$ measure.   
\bigskip
\\
{\bf Insert Figure 5 here.}   
\bigskip

\begin{figure}[H]
\centering
{\includegraphics[width=16cm]{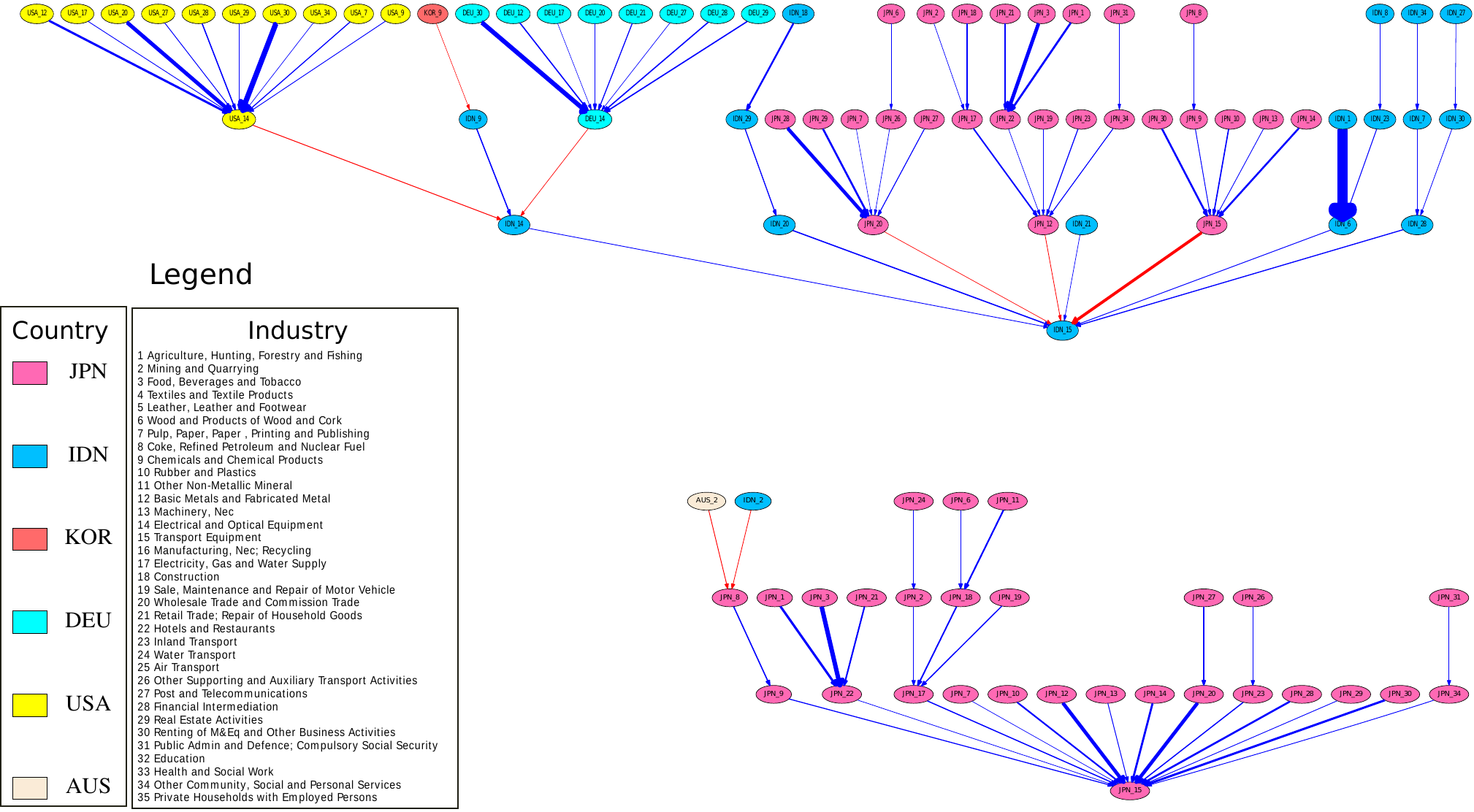}}
\caption{{\bf The comparison of the transport equipment industry between Indonesia and Japan in 1995.} The upper and lower GVTs are rooted in the tranport equipment industry in Indonesia and Japan respectively in 1995. The edge weight threshold is set to 0.01 and the number of rounds is limited to 3. Different colors of the nodes indicate different countries. The red edges indicate cross-country relationships while the blue edges indicate domestic relationships. The edge width is proportional to the edge weight, i.e., the share of the value-added contribution.} \label{fig_4}
\end{figure}

We can also examine the different structures of the GVTs for the same industry and the same year but for different countries. Figure \ref{fig_4} compares the transport equipment industry between Indonesia and Japan in 1995. The immediate conclusion from this comparison is that the transport equipment industry has a more international GVT in Indonesia than in Japan. More interestingly, Japan's industries actually play important roles in Indonesia's GVT, i.e., three Japan's industries (JPN\_12, JPN\_15, and JPN\_20) are direct neighbors of the root in Indonesia (This observation coincides with the increased foreign direct investment from Japan to Indonesia's car industry in 1995.). In this simple comparison, JPN\_15 is clearly more competitive than IDN\_15, according to the above $TI$ measure. 
\bigskip
\\
{\bf Insert Figure 6 here.}   
\bigskip

In summary, previous studies of the GVCs are mainly interested in knowing how global the GVCs are rather than how the GVCs look like. To fill the gap in the literature, our paper is the first attempt to investigate the topological properties of the industry-level GVCs. From a complex networks perspective, we map the World Input-Output Database (WIOD) into the global value networks (GVNs), where the nodes are the individual industries in different countries and the edges are the value-added contribution relationships. 

Based on the GVNs, the global value trees (GVTs) can be obtained by a breadth-first search algorithm with a threshold of edge weight and a limit of the number of rounds. We compute the GVTs for all the industries available in the WIOD and explore some basic properties of the GVTs. In particular, we estimate the allometric scaling exponents and find that the GVTs are topologically more similar to a star than to a chain. However, the GVTs have become more and more hierarchical over time. We also develop an industry importance measure based on the GVTs and compare it with other network centrality measures of the industries. We find that the tree-based measure performs the best in terms of the correlation with the industry total value-added. Therefore, the GVTs still retain the essential information of the GVNs and can be viewed as a reasonable simplification of the latter. Finally, we discuss some future applications of the GVTs such as to examine the evolution of the GVTs for a certain industry and to compare the GVTs of the same industry in different countries. 

\section*{Acknowledgments}

Authors acknowledge insightful discussions with Guido Caldarelli. M.R. and Z.Z. acknowledge funding from the Ministry of Education, Universities and Research (MIUR) - the Basic Research Investment Fund (FIRB) project RBFR12BA3Y and from the ‘ViWaN: The Global Virtual Water Network’ project. All authors acknowledge support from the Future Emerging Technologies (FET) projects MULTIPLEX 317532 and SIMPOL 610704 and the National Research Project (PNR) project CRISIS Lab. F.C. acknowledges Sardinia Regional Government for the financial support of her PhD scholarship (P.O.R. Sardegna F.S.E. Operational Programme of the Autonomous Region of Sardinia, European Social Fund 2007–2013 - Axis IV Human Resources, Objective l.3, Line of Activity l.3.1.).

\section*{Author Contributions}

A.C., M.P., M.R., and Z.Z. conceived and designed the research. F.C., M.P., and Z.Z. performed the computation. M.R. and Z.Z. analyzed the results. Z.Z. wrote the main text. 

\section*{Additional Information}
The authors declare no competing financial interests.



\newpage \clearpage

\section*{Figures}

\section*{Tables}

\begin{table}[H]
\caption{{\bf A hypothetical two-economy-two-industry MRIO table.} The $4\times4$ inter-industry transactions matrix records outputs selling in its rows and inputs buying in its columns. The additional columns are the final demand and the additional row is the value added. Finally, the last column and the last row record the total industry outputs.}
  \centering
  \resizebox{16cm}{!}{
    \begin{tabular}{@{\extracolsep{4pt}}ccccccccc@{}}
    \toprule
          &       & \multicolumn{4}{c}{Buyer Industry} &       &       &  \\
        \cline{3-6}
          &       & \multicolumn{2}{c}{Economy 1} & \multicolumn{2}{c}{Economy 2} & \multicolumn{2}{c}{Final Demand} &  \\
          \cline{3-4} \cline{5-6} \cline{7-8}
          \multicolumn{2}{c}{Seller Industry}        & Industry 1 & Industry 2 & Industry 1 & Industry 2 & Economy 1 & Economy 2 & Total Output \\
          \colrule
     \multirow{2}{*}{Economy 1} & Industry 1 & 25    & 10     & 20    & 10    & 45     & 10     & 120 \\
      & Industry 2 & 10     & 5    & 10    & 20    & 50     & 5     & 100 \\
    \multirow{2}{*}{Economy 2} & Industry 1 & 30     & 15     & 600  & 500   & 5  & 8650  & 9800 \\
     & Industry 2 & 35     & 30     & 1000   & 1000   & 25  & 7910  & 10000 \\
      \multicolumn{2}{c}{Value Added}  & 20    & 40    & 8170  & 8470  &       &       &  \\
     \multicolumn{2}{c}{Total Output} & 120    & 100    & 9800 & 10000 &       &       &  \\
          \botrule
    \end{tabular}%
    }
  \label{table_1}%
\end{table} 

\begin{table}[H]
  \centering
  \caption{{\bf The summary statistics of the size of the GVTs by varying the weight threshold, i.e., $\alpha$, for the selected years.} The number of observations is the number of nonempty GVTs according to the value of $\alpha$. CV stands for the coefficient of variation, which is the standard deviation divided by the mean.}
  \resizebox{12cm}{!}{
    \begin{tabular}{@{\extracolsep{4pt}}cccccccccc@{}}
    \toprule
     & \multicolumn{3}{c}{$\alpha=0.1$} & \multicolumn{3}{c}{$\alpha=0.01$} & \multicolumn{3}{c}{$\alpha=0.001$} \\
    \cline{2-4} \cline{5-7} \cline{8-10}
     & \textbf{1995} & \textbf{2003} & \textbf{2011} & \textbf{1995} & \textbf{2003} & \textbf{2011} & \textbf{1995} & \textbf{2003} & \textbf{2011} \\
    \textbf{\# Obs.} & 182   & 184   & 194   & 1344  & 1345  & 1342  & 1348  & 1347  & 1345 \\
    \textbf{Min} & 2     & 2     & 2     & 9     & 2     & 2     & 196   & 173   & 159 \\
    \textbf{Max} & 4     & 4     & 5     & 149   & 221   & 172   & 1165  & 1150  & 1162 \\
    \textbf{Mean} & 2.104 & 2.076 & 2.062 & 48.089 & 43.888 & 44.242 & 793.677 & 856.200 & 855.436 \\
    \textbf{CV} & 0.162 & 0.147 & 0.145 & 0.489 & 0.527 & 0.531 & 0.194 & 0.198 & 0.197 \\
    \botrule
    \end{tabular}%
    }
  \label{table_2}%
\end{table}%

\begin{table}[H]
  \centering
  \caption{{\bf The top-20 industries identified by the tree-based importance measure and other network centrality measures for the selected years.} $TI$ is the tree-based importance measure, $CC$ is the closeness centrality, $BC$ is the betweenness centrality, $PR$ is the PageRank centrality, $VT$ is the industry total value-added. The full names of the corresponding industries of the 3-letter codes can be found in Table S2 in Supplementary Information.}
  \resizebox{16cm}{!}{
    \begin{tabular}{@{\extracolsep{4pt}}cccccccccccccccc@{}}
    \toprule
          & \multicolumn{5}{c}{\textbf{1995}}     & \multicolumn{5}{c}{\textbf{2003}}     & \multicolumn{5}{c}{\textbf{2011}} \\
    \cline{2-6} \cline{7-11} \cline{12-16}
    \textbf{Rank} & $TI$    & $CC$    & $BC$    & $PR$    & $VT$    & $TI$    & $CC$    & $BC$    & $PR$    & $VT$    & $TI$    & $CC$    & $BC$    & $PR$    & $VT$ \\
    \textbf{1} & USA-Pup & DEU-Tpt & USA-Obs & USA-Pub & USA-Pub & USA-Cst & USA-Obs & USA-Obs & USA-Pub & USA-Obs & USA-Obs & CHN-Cst & CHN-Cst & USA-Pub & USA-Obs \\
    \textbf{2} & CAN-Pup & DEU-Obs & DEU-Obs & USA-Cst & USA-Est & USA-Obs & USA-Tpt & DEU-Tpt & USA-Hth & USA-Pub & USA-Htl & CHN-Met & USA-Pub & CHN-Cst & USA-Pub \\
    \textbf{3} & USA-Cst & DEU-Est & USA-Pub & USA-Hth & USA-Obs & USA-Est & USA-Pub & USA-Tpt & USA-Cst & USA-Est & USA-Cst & CHN-Omn & USA-Obs & USA-Hth & USA-Est \\
    \textbf{4} & USA-Obs & DEU-Cst & USA-Elc & JPN-Cst & USA-Rtl & USA-Htl & USA-Hth & DEU-Obs & USA-Est & USA-Fin & CHN-Elc & CHN-Min & RUS-Min & CHN-Pub & USA-Fin \\
    \textbf{5} & JPN-Ele & DEU-Elc & DEU-Tpt & USA-Est & USA-Fin & USA-Ele & USA-Fin & USA-Pub & USA-Tpt & USA-Rtl & AUS-Min & CHN-Whl & DEU-Obs & DEU-Tpt & USA-Hth \\
    \textbf{6} & USA-Est & DEU-Fod & RUS-Min & USA-Tpt & JPN-Est & USA-Cok & USA-Cst & DEU-Cok & CHN-Cst & USA-Hth & RUS-Min & CHN-Elc & DEU-Tpt & CHN-Elc & USA-Rtl \\
    \textbf{7} & GBR-Fin & USA-Obs & FRA-Obs & USA-Htl & USA-Whl & USA-Wod & USA-Est & JPN-Tpt & DEU-Tpt & USA-Whl & USA-Ocm & CHN-Fin & CHN-Elc & USA-Cst & USA-Whl \\
    \textbf{8} & DEU-Fin & USA-Pub & JPN-Elc & USA-Rtl & USA-Hth & GBR-Fin & USA-Rtl & DEU-Elc & USA-Htl & USA-Cst & JPN-Ele & CHN-Ldt & CHN-Agr & CHN-Hth & JPN-Est \\
    \textbf{9} & JPN-Elc & FRA-Obs & USA-Fod & DEU-Cst & JPN-Whl & USA-Pup & USA-Whl & USA-Elc & USA-Rtl & JPN-Est & USA-Met & CHN-Agr & GBR-Obs & USA-Htl & CHN-Agr \\
    \textbf{10} & USA-Elc & DEU-Hth & DEU-Met & USA-Fod & JPN-Cst & USA-Ldt & USA-Ocm & USA-Fod & USA-Ocm & USA-Ocm & USA-Otr & CHN-Ele & ESP-Obs & USA-Tpt & USA-Ocm \\
    \textbf{11} & USA-Ele & DEU-Met & USA-Tpt & DEU-Tpt & USA-Cst & USA-Chm & USA-Htl & JPN-Whl & USA-Fod & USA-Htl & USA-Whl & CHN-Chm & FIN-Obs & USA-Rtl & USA-Cst \\
    \textbf{12} & USA-Ldt & USA-Elc & JPN-Whl & DEU-Fod & JPN-Fin & USA-Met & USA-Fod & RUS-Min & GBR-Pub & JPN-Pub & USA-Cok & CHN-Fod & USA-Fin & USA-Est & CHN-Whl \\
    \textbf{13} & JPN-Cst & USA-Whl & JPN-Cst & USA-Ocm & JPN-Pub & CAN-Min & USA-Elc & JPN-Cst & JPN-Cst & JPN-Whl & USA-Ele & CHN-Obs & USA-Cok & CHN-Mch & JPN-Pub \\
    \textbf{14} & JPN-Sal & USA-Tpt & DEU-Elc & JPN-Pub & JPN-Obs & DEU-Fin & USA-Pst & CHN-Elc & DEU-Fod & JPN-Obs & USA-Min & CHN-Pst & CHN-Tex & CHN-Tpt & CHN-Cst \\
    \textbf{15} & JPN-Ldt & USA-Hth & DEU-Tex & DEU-Hth & JPN-Rtl & USA-Otr & DEU-Tpt & DEU-Met & DEU-Cst & USA-Pst & USA-Fin & CHN-Mch & DEU-Met & USA-Fod & USA-Htl \\
    \textbf{16} & USA-Chm & USA-Cst & DEU-Cok & GBR-Pub & USA-Ocm & JPN-Ele & USA-Met & ESP-Tpt & CHN-Pub & DEU-Obs & CAN-Min & CHN-Otr & FIN-Cst & USA-Ocm & JPN-Obs \\
    \textbf{17} & USA-Htl & USA-Fin & FRA-Ele & DEU-Htl & DEU-Est & USA-Whl & USA-Agr & FRA-Obs & GBR-Est & JPN-Cst & CHN-Obs & USA-Obs & TUR-Tex & GBR-Hth & DEU-Obs \\
    \textbf{18} & JPN-Cok & USA-Rtl & DEU-Cst & DEU-Est & DEU-Obs & CHN-Elc & USA-Pup & ITA-Obs & FRA-Tpt & DEU-Est & USA-Pub & USA-Pub & FRA-Obs & CHN-Fod & JPN-Whl \\
    \textbf{19} & DEU-Cst & USA-Est & JPN-Met & CHN-Cst & JPN-Ocm & DEU-Met & USA-Chm & CHN-Cst & USA-Fin & JPN-Fin & GBR-Fin & USA-Hth & AUS-Min & CHN-Edu & CHN-Est \\
    \textbf{20} & JPN-Htl & DEU-Fin & DEU-Fod & FRA-Tpt & USA-Pst & JPN-Sal & DEU-Obs & USA-Agr & ESP-Cst & GBR-Obs & USA-Fod & CHN-Wod & IDN-Min & JPN-Cst & CHN-Met \\
    \botrule
    \end{tabular}%
    }
  \label{table_3}%
\end{table}%

\begin{table}[H]
  \centering
  \caption{{\bf The Pearson correlation coefficient matrix between the tree-based importance measure and other network centrality measures (in logarithm) for the selected years.} The size of the sample is in the parentheses next to the corresponding years. $TI$ is the tree-based importance measure, $CC$ is the closeness centrality, $BC$ is the betweenness centrality, $PR$ is the PageRank centrality, $VT$ is the industry total value-added. $\ast\ast$ means that the coefficient is significant at 1\% level. $\ast$ means that the coefficient is significant at 5\% level.}
  \resizebox{16cm}{!}{
    \begin{tabular}{@{\extracolsep{4pt}}cccccccccccccccccc@{}}
    \toprule
    \multicolumn{6}{c}{\textbf{1995} (\# Obs. 384)}        & \multicolumn{6}{c}{\textbf{2003} (\# Obs. 351)}        & \multicolumn{6}{c}{\textbf{2011} (\# Obs. 324)} \\
    \cline{1-6} \cline{7-12} \cline{13-18}
         & $log(TI)$ & $log(CC)$ & $log(BC)$ & $log(PR)$ & $log(VT)$ &       & $log(TI)$ & $log(CC)$ & $log(BC)$ & $log(PR)$ & $log(VT)$ &       & $log(TI)$ & $log(CC)$ & $log(BC)$ & $log(PR)$ & $log(VT)$ \\
    $log(TI)$ & 1     & -     & -     & -     & -     & $log(TI)$ & 1     & -     & -     & -     & -     & $log(TI)$ & 1     & -     & -     & -     & - \\
    $log(CC)$ & 0.577$\ast\ast$ & 1     & -     & -     & -     & $log(CC)$ & 0.560$\ast\ast$ & 1     & -     & -     & -     & $log(CC)$ & 0.523$\ast\ast$ & 1     & -     & -     & - \\
    $log(BC)$ & 0.282$\ast\ast$ & 0.196$\ast\ast$ & 1     & -     & -     & $log(BC)$ & 0.229$\ast\ast$ & 0.063 & 1     & -     & -     & $log(BC)$ & 0.276$\ast\ast$ & 0.082 & 1     & -     & - \\
    $log(PR)$ & 0.436$\ast\ast$ & 0.363$\ast\ast$ & 0.370$\ast\ast$ & 1     & -     & $log(PR)$ & 0.450$\ast\ast$ & 0.395$\ast\ast$ & 0.317$\ast\ast$ & 1     & -     & $log(PR)$ & 0.388$\ast\ast$ & 0.328$\ast\ast$ & 0.235$\ast\ast$ & 1     & - \\
    $log(VT)$ & 0.838$\ast\ast$ & 0.703$\ast\ast$ & 0.325$\ast\ast$ & 0.663$\ast\ast$ & 1     & $log(VT)$ & 0.814$\ast\ast$ & 0.727$\ast\ast$ & 0.256$\ast\ast$ & 0.684$\ast\ast$ & 1     & $log(VT)$ & 0.820$\ast\ast$ & 0.639$\ast\ast$ & 0.276$\ast\ast$ & 0.623$\ast\ast$ & 1 \\
    \botrule
    \end{tabular}%
    }
  \label{table_4}%
\end{table}%

\section*{Supplementary Information}

\setcounter{table}{0}
\renewcommand\thetable{S\arabic{table}}

\setcounter{figure}{0}
\renewcommand\thefigure{S\arabic{figure}}


\begin{table}[H] 
  \centering
  \caption{{\bf The list of WIOD economies.}}
  \resizebox{16cm}{!}{
    \begin{tabular}{@{\extracolsep{4pt}}llllllllll@{}}
    \toprule
    \multicolumn{2}{c}{\textbf{Euro-Zone}} & \multicolumn{2}{c}{\textbf{Non-Euro EU}} & \multicolumn{2}{c}{\textbf{NAFTA}} & \multicolumn{2}{c}{\textbf{East Asia}} & \multicolumn{2}{c}{\textbf{BRIIAT}} \\
    \cline{1-2} \cline{3-4} \cline{5-6}  \cline{7-8} \cline{9-10}
    \textbf{Economy} & \textbf{3L Code} & \textbf{Economy} & \textbf{3L Code} & \textbf{Economy} & \textbf{3L Code} & \textbf{Economy} & \textbf{3L Code} & \textbf{Economy} & \textbf{3L Code} \\
    
    Austria & AUT   & Bulgaria & BGR   & Canada & CAN   & China & CHN   & Australia & AUS \\
    Belgium & BEL   & Czech Rep. & CZE   & Mexico & MEX   & Japan & JPN   & Brazil & BRA \\
    Cyprus & CYP   & Denmark & DNK   & USA   & USA   & South Korea & KOR   & India & IND \\
    Estonia & EST   & Hungary & HUN   &       &       & Taiwan & TWN   & Indonesia & IDN \\
    Finland & FIN   & Latvia & LVA   &       &       &       &       & Russia & RUS \\
    France & FRA   & Lithuania & LTU   &       &       &       &       & Turkey & TUR \\
    Germany & DEU   & Poland & POL   &       &       &       &       &       &  \\
    Greece & GRC   & Romania & ROM   &       &       &       &       &       &  \\
    Ireland & IRL   & Sweden & SWE   &       &       &       &       &       &  \\
    Italy & ITA   & UK    & GBR   &       &       &       &       &       &  \\
    Luxembourg & LUX   &       &       &       &       &       &       &       &  \\
    Malta & MLT   &       &       &       &       &       &       &       &  \\
    Netherlands & NLD   &       &       &       &       &       &       &       &  \\
    Portugal & PRT   &       &       &       &       &       &       &       &  \\
    Slovakia & SVK   &       &       &       &       &       &       &       &  \\
    Slovenia & SVN   &       &       &       &       &       &       &       &  \\
    Spain & ESP   &       &       &       &       &       &       &       &  \\
    \botrule
    \end{tabular}%
    }
  \label{table_A1}%
\end{table}%

\newpage 

\begin{table}
  \centering
  \caption{{\bf The list of WIOD industries.}}
  \resizebox{16cm}{!}{
    \begin{tabular}{@{\extracolsep{4pt}}llll@{}}
    \toprule
    \textbf{Full Name} & \textbf{ISIC Rev. 3 Code} & \textbf{WIOD Code} & \textbf{3-Letter Code} \\
    \cline{1-1} \cline{2-2} \cline{3-3} \cline{4-4}
    Agriculture, Hunting, Forestry and Fishing & AtB   & c1    & Agr \\
    Mining and Quarrying & C     & c2    & Min \\
    Food, Beverages and Tobacco & 15t16 & c3    & Fod \\
    Textiles and Textile Products & 17t18 & c4    & Tex \\
    Leather, Leather and Footwear & 19    & c5    & Lth \\
    Wood and Products of Wood and Cork & 20    & c6    & Wod \\
    Pulp, Paper, Paper , Printing and Publishing & 21t22 & c7    & Pup \\
    Coke, Refined Petroleum and Nuclear Fuel & 23    & c8    & Cok \\
    Chemicals and Chemical Products & 24    & c9    & Chm \\
    Rubber and Plastics & 25    & c10   & Rub \\
    Other Non-Metallic Mineral & 26    & c11   & Omn \\
    Basic Metals and Fabricated Metal & 27t28 & c12   & Met \\
    Machinery, Nec & 29    & c13   & Mch \\
    Electrical and Optical Equipment & 30t33 & c14   & Elc \\
    Transport Equipment & 34t35 & c15   & Tpt \\
    Manufacturing, Nec; Recycling & 36t37 & c16   & Mnf \\
    Electricity, Gas and Water Supply & E     & c17   & Ele \\
    Construction & F     & c18   & Cst \\
    Sale, Maintenance and Repair of Motor Vehicles and Motorcycles; Retail Sale of Fuel & 50    & c19   & Sal \\
    Wholesale Trade and Commission Trade, Except of Motor Vehicles and Motorcycles & 51    & c20   & Whl \\
    Retail Trade, Except of Motor Vehicles and Motorcycles; Repair of Household Goods & 52    & c21   & Rtl \\
    Hotels and Restaurants & H     & c22   & Htl \\
    Inland Transport & 60    & c23   & Ldt \\
    Water Transport & 61    & c24   & Wtt \\
    Air Transport & 62    & c25   & Ait \\
    Other Supporting and Auxiliary Transport Activities; Activities of Travel Agencies & 63    & c26   & Otr \\
    Post and Telecommunications & 64    & c27   & Pst \\
    Financial Intermediation & J     & c28   & Fin \\
    Real Estate Activities & 70    & c29   & Est \\
    Renting of M\&Eq and Other Business Activities & 71t74 & c30   & Obs \\
    Public Admin and Defence; Compulsory Social Security & L     & c31   & Pub \\
    Education & M     & c32   & Edu \\
    Health and Social Work & N     & c33   & Hth \\
    Other Community, Social and Personal Services & O     & c34   & Ocm \\
    Private Households with Employed Persons & P     & c35   & Pvt \\
    \botrule
    \end{tabular}%
    }
  \label{table_A2}%
\end{table}%

\begin{table}[!h]
  \centering
  \caption{{\bf The country rankings based on the tree-based importance measure and other network centrality measures for the selected years.} $TI$ is the tree-based importance measure, $CC$ is the closeness centrality, $BC$ is the betweenness centrality, $PR$ is the PageRank centrality, $VT$ is the industry total value-added. The full names of the corresponding industries of the 3-letter codes can be found in Table \ref{table_A2} in Supplementary Information.}
  \resizebox{16cm}{!}{
    \begin{tabular}{@{\extracolsep{16pt}}cccccccccccccccc@{}}
    \toprule
          & \multicolumn{5}{c}{\textbf{1995}}     & \multicolumn{5}{c}{\textbf{2003}}     & \multicolumn{5}{c}{\textbf{2011}} \\
    \cline{2-6} \cline{7-11} \cline{12-16}
    \textbf{Rank} & $TI$    & $CC$    & $BC$    & $PR$    & $VT$    & $TI$    & $CC$    & $BC$    & $PR$    & $VT$    & $TI$    & $CC$    & $BC$    & $PR$    & $VT$ \\
    \textbf{1} & \textbf{USA} & \textbf{USA} & \textbf{DEU} & \textbf{USA} & \textbf{USA} & \textbf{USA} & \textbf{USA} & \textbf{USA} & \textbf{USA} & \textbf{USA} & \textbf{USA} & \textbf{USA} & \textbf{CHN} & \textbf{USA} & \textbf{USA} \\
    \textbf{2} & \textbf{JPN} & \textbf{FRA} & \textbf{USA} & \textbf{DEU} & \textbf{JPN} & \textbf{JPN} & \textbf{FRA} & \textbf{DEU} & \textbf{DEU} & \textbf{JPN} & \textbf{CHN} & \textbf{FRA} & \textbf{USA} & \textbf{CHN} & \textbf{CHN} \\
    \textbf{3} & \textbf{DEU} & \textbf{CAN} & \textbf{JPN} & \textbf{JPN} & \textbf{DEU} & \textbf{DEU} & \textbf{MEX} & \textbf{JPN} & \textbf{GBR} & \textbf{DEU} & \textbf{JPN} & \textbf{CAN} & \textbf{DEU} & \textbf{DEU} & \textbf{JPN} \\
    \textbf{4} & \textbf{GBR} & \textbf{MEX} & \textbf{FRA} & \textbf{GBR} & \textbf{FRA} & \textbf{GBR} & \textbf{CAN} & \textbf{CHN} & \textbf{CHN} & \textbf{GBR} & \textbf{AUS} & \textbf{MEX} & \textbf{RUS} & \textbf{GBR} & \textbf{DEU} \\
    \textbf{5} & \textbf{CAN} & \textbf{GBR} & \textbf{RUS} & \textbf{ITA} & \textbf{GBR} & \textbf{FRA} & \textbf{GBR} & \textbf{ITA} & \textbf{ITA} & \textbf{FRA} & \textbf{DEU} & \textbf{GBR} & \textbf{ESP} & \textbf{ITA} & \textbf{FRA} \\
    \textbf{6} & \textbf{FRA} & \textbf{TWN} & \textbf{ITA} & \textbf{FRA} & \textbf{ITA} & \textbf{CAN} & \textbf{DEU} & \textbf{ESP} & \textbf{JPN} & \textbf{CHN} & \textbf{RUS} & \textbf{JPN} & \textbf{GBR} & \textbf{FRA} & \textbf{GBR} \\
    \textbf{7} & \textbf{ITA} & \textbf{DEU} & \textbf{GBR} & \textbf{ESP} & \textbf{CHN} & \textbf{CHN} & \textbf{JPN} & \textbf{RUS} & \textbf{FRA} & \textbf{ITA} & \textbf{CAN} & \textbf{DEU} & \textbf{FIN} & \textbf{JPN} & \textbf{BRA} \\
    \textbf{8} & \textbf{BRA} & \textbf{JPN} & \textbf{ESP} & \textbf{KOR} & \textbf{BRA} & \textbf{ITA} & \textbf{ITA} & \textbf{FRA} & \textbf{ESP} & \textbf{ESP} & \textbf{GBR} & \textbf{BRA} & \textbf{KOR} & \textbf{RUS} & \textbf{ITA} \\
    \textbf{9} & \textbf{AUS} & \textbf{ITA} & \textbf{FIN} & \textbf{CHN} & \textbf{ESP} & \textbf{RUS} & \textbf{ESP} & \textbf{FIN} & \textbf{KOR} & \textbf{CAN} & \textbf{FRA} & \textbf{ITA} & \textbf{AUS} & \textbf{ESP} & \textbf{IND} \\
    \textbf{10} & \textbf{RUS} & \textbf{KOR} & \textbf{AUS} & \textbf{GRC} & \textbf{CAN} & \textbf{MEX} & \textbf{KOR} & \textbf{GBR} & \textbf{TUR} & \textbf{MEX} & \textbf{ITA} & \textbf{RUS} & \textbf{BRA} & \textbf{BRA} & \textbf{CAN} \\
    11    & ESP   & ESP   & BRA   & BRA   & KOR   & ESP   & IND   & IND   & CAN   & KOR   & BRA   & IND   & JPN   & CAN   & ESP \\
    12    & CHN   & AUS   & ROM   & RUS   & NLD   & AUS   & AUS   & TUR   & MEX   & IND   & IND   & AUS   & FRA   & IND   & AUS \\
    13    & KOR   & BRA   & DNK   & AUS   & IND   & KOR   & NLD   & AUS   & GRC   & BRA   & KOR   & KOR   & CAN   & TUR   & RUS \\
    14    & IND   & NLD   & IDN   & TUR   & AUS   & IND   & BRA   & IDN   & AUS   & AUS   & MEX   & ESP   & ITA   & KOR   & MEX \\
    15    & IDN   & RUS   & KOR   & NLD   & RUS   & BRA   & TWN   & BRA   & RUS   & NLD   & ESP   & TUR   & IDN   & AUS   & KOR \\
    16    & NLD   & IND   & BEL   & IND   & MEX   & NLD   & RUS   & DNK   & IND   & RUS   & TWN   & TWN   & TUR   & GRC   & IDN \\
    17    & MEX   & BEL   & IND   & CAN   & BEL   & TWN   & BEL   & POL   & NLD   & TWN   & IDN   & NLD   & MEX   & MEX   & NLD \\
    18    & SWE   & DNK   & TUR   & SWE   & TWN   & IDN   & TUR   & HUN   & SWE   & BEL   & TUR   & BEL   & GRC   & POL   & TUR \\
    19    & BEL   & SWE   & CHN   & AUT   & IDN   & TUR   & SWE   & BGR   & BRA   & SWE   & NLD   & POL   & LVA   & IDN   & SWE \\
    20    & TUR   & POL   & MEX   & BEL   & SWE   & SWE   & POL   & ROM   & POL   & TUR   & SWE   & FIN   & CZE   & NLD   & BEL \\
    21    & TWN   & FIN   & BGR   & DNK   & AUT   & BEL   & GRC   & CAN   & AUT   & IDN   & POL   & GRC   & HUN   & SWE   & POL \\
    22    & AUT   & AUT   & GRC   & PRT   & TUR   & DNK   & DNK   & KOR   & BEL   & AUT   & BEL   & PRT   & POL   & BEL   & TWN \\
    23    & DNK   & PRT   & NLD   & TWN   & DNK   & AUT   & PRT   & SVK   & IDN   & POL   & DNK   & AUT   & IND   & FIN   & AUT \\
    24    & GRC   & GRC   & CAN   & MEX   & POL   & POL   & AUT   & GRC   & FIN   & DNK   & AUT   & DNK   & SVK   & AUT   & DNK \\
    25    & POL   & CZE   & LVA   & FIN   & GRC   & GRC   & FIN   & TWN   & DNK   & GRC   & GRC   & IRL   & CYP   & ROM   & GRC \\
    26    & FIN   & IRL   & HUN   & IDN   & FIN   & FIN   & IRL   & PRT   & PRT   & FIN   & PRT   & ROM   & SWE   & DNK   & FIN \\
    27    & PRT   & CHN   & PRT   & POL   & PRT   & PRT   & CHN   & EST   & IRL   & IRL   & FIN   & CZE   & ROM   & PRT   & PRT \\
    28    & CZE   & IDN   & POL   & CZE   & IRL   & IRL   & CZE   & BEL   & CZE   & PRT   & ROM   & HUN   & PRT   & CZE   & IRL \\
    29    & ROM   & ROM   & SWE   & HUN   & CZE   & CZE   & IDN   & CZE   & HUN   & CZE   & CZE   & SVK   & DNK   & CYP   & CZE \\
    30    & HUN   & TUR   & TWN   & ROM   & HUN   & HUN   & HUN   & LVA   & CYP   & HUN   & HUN   & BGR   & SVN   & HUN   & ROM \\
    31    & IRL   & HUN   & AUT   & CYP   & ROM   & ROM   & ROM   & SWE   & TWN   & ROM   & IRL   & CHN   & BEL   & TWN   & HUN \\
    32    & SVN   & SVN   & CYP   & IRL   & LUX   & SVN   & SVK   & CYP   & ROM   & SVK   & SVK   & IDN   & BGR   & IRL   & SVK \\
    33    & SVK   & BGR   & CZE   & SVN   & SVN   & SVK   & SVN   & MEX   & SVN   & LUX   & SVN   & SWE   & MLT   & BGR   & LUX \\
    34    & BGR   & SVK   & LTU   & BGR   & SVK   & BGR   & LTU   & LTU   & LTU   & SVN   & BGR   & LTU   & LUX   & SVK   & BGR \\
    35    & CYP   & LUX   & SVN   & LVA   & BGR   & LTU   & BGR   & MLT   & LVA   & BGR   & CYP   & SVN   & LTU   & LVA   & SVN \\
    36    & LUX   & LTU   & IRL   & LTU   & CYP   & CYP   & LUX   & NLD   & SVK   & LTU   & LTU   & EST   & EST   & LTU   & LTU \\
    37    & LTU   & CYP   & EST   & SVK   & LTU   & LUX   & CYP   & SVN   & BGR   & CYP   & LVA   & LVA   & TWN   & SVN   & LVA \\
    38    & LVA   & EST   & SVK   & EST   & LVA   & LVA   & LVA   & AUT   & EST   & LVA   & EST   & LUX   & NLD   & EST   & CYP \\
    39    & EST   & MLT   & MLT   & MLT   & EST   & EST   & EST   & IRL   & MLT   & EST   & LUX   & CYP   & AUT   & MLT   & EST \\
    40    & MLT   & LVA   & LUX   & LUX   & MLT   & MLT   & MLT   & LUX   & LUX   & MLT   & MLT   & MLT   & IRL   & LUX   & MLT \\
    \botrule
    \end{tabular}%
    }
  \label{table_A3}%
\end{table}%

\end{document}